\documentclass[aps,prl,twocolumn,amsmath,amssymb,nobibnotes]{revtex4-2}
\usepackage{bm} 
\usepackage{graphicx}
\usepackage{dcolumn}
\usepackage{gensymb}
\usepackage{hyperref}
\hypersetup{
    colorlinks=true,
    linkcolor=blue,
    filecolor=blue,      
    urlcolor=blue,
    citecolor= blue,}

\usepackage{bm}

\begin{document}

\title{Disentangling the influence of excitation energy and compound nucleus angular momentum on fission fragment angular momentum}
\author{Simone Cannarozzo}
 \email[]{simone.cannarozzo@physics.uu.se}
\author{Stephan Pomp}
 \email[]{stephan.pomp@physics.uu.se}
\author{Andreas Solders}
\author{Ali Al-Adili}
\author{Zhihao Gao}
\author{Mattias Lantz}
\affiliation{Department of Physics and Astronomy, Uppsala University, Box 516, Uppsala, 751 20, Sweden}

\author{Heikki Penttil{\"a}}
\author{Anu Kankainen}
\author{Iain Moore}
\author{Tommi Eronen}
\author{Zhuang Ge}
\author{Jouni Ruotsalainen}
\author{Maxime Mougeot}
\author{Ville Virtanen}
\author{Arthur Jaries}
\author{Marek Stryjczyk}
\author{Andrea Raggio}
\affiliation{Department of Physics, Accelerator Laboratory, University of Jyvaskyla, P.O. Box 35(YFL), Jyvaskyla, 40014, Finland}

\begin{abstract}
The origin of the large angular momenta observed for fission fragments is still a question under discussion.
To address this, we study isomeric yield ratios (IYR), \textit{i.e.} the relative population of two or more long-lived metastable states with different spins, of fission products.

We report on IYR of 17 isotopes produced in the 28 MeV $\alpha$-induced fission of $^{232}$Th at the IGISOL facility of the University of Jyv{\"a}skyl{\"a}. 
The fissioning nuclei in this reaction are $^{233,234,235}$U*.
We compare our data to IYR from thermal neutron-induced fission of $^{233}$U and $^{235}$U, and we observe statistically significant larger IYR in the $^{232}$Th($\alpha$,f) reaction, where the average compound nucleus (CN) spin is 7.7 $\hbar$, than in $^{233,235}$U(n$_{th}$,f), with average spins 2.6 and 3.6 $\hbar$, respectively.

To assess the influence of the excitation energy, we study literature data of IYR from photon-induced fission reactions, and find that, within current uncertainties, the IYR indicate no dependency of the CN excitation energy. 
We conclude that the different IYR seem to be due to the different CN spins alone.
This would imply that the FF angular momentum only partly comes from the fission process itself, and is in addition influenced by the angular momentum present in the CN.
\end{abstract}

\maketitle

Since its discovery in 1939 \cite{Hahn1939}, considerable progress has been made in advancing our understanding of nuclear fission.  
However, several fundamental questions remain unresolved 
\cite{andreyevNuclearFissionReview2018,Schmidt2018}. 
One concerns the origin of fission fragment angular momenta, first raised by Wilhelmy \textit{et al.} \cite{Wilhelmy}.

In binary fission, after the nucleus reaches the saddle point, two clusters of nucleons begin to form the two nascent fragments. 
Wilhelmy \textit{et al.} observed that the fragments emerging after scission typically carry significant angular momentum (in the order of 7 $\hbar$ \cite{Wilhelmy}),
even in spontaneous fission of the 0$^+$ spin nucleus $^{252}$Cf. 
The physical process responsible for the generation of fission fragment (FF) spin is still highly debated \cite{andreyevNuclearFissionReview2018,wilsonAngularMomentumGeneration2021,randrupGenerationFragmentAngular2021,marevicAngularMomentumFission2021,vogtAngularMomentumEffects2021,bulgacAngularCorrelationFission2022,Scamps2023}.

The two most widely accepted explanations suggest that the angular momentum can be partially determined by the excitation of collective relative orbital motions between the two prefragments before scission \cite{RASMUSSEN1969465,MORETTO1989453,BulgacAMmodes,Dossing}, and/or by Coulomb repulsion around scission \cite{HoffmanCoulomb,RandrupCoulomb}. 
At the same time, other factors may also influence the angular momentum generation, such as the deformation of the FFs and the nuclear shell structure \cite{marevicAngularMomentumFission2021,WILKINSdeformedshell,Bertschdeformation}. 

Recently, Wilson \textit{et al.} proposed post-scission generation of the angular momentum \cite{wilsonAngularMomentumGeneration2021}. They experimentally observed the same saw-tooth shaped dependency of the FFs angular momentum on the mass for three different low-energy fissioning systems, concluding that completely uncorrelated torques are generated by the rupture of the neck connecting the nascent fragments \cite{wilsonAngularMomentumGeneration2021}.
However, uncorrelated angular momenta have later also been obtained through pre-scission model calculations by Randrup \textit{et al.} \cite{randrupGenerationFragmentAngular2021} and Scamps \textit{et al.} \cite{Scamps2023}, showing that the post-scission model advanced by Wilson \textit{et al.} is not the only possible explanation for their observation. 

Any model describing angular momenta of FFs must account for the influence of both excitation energy and angular momentum of the compound nucleus (CN). 
The influence of CN conditions on FF spins has been the focus of extensive research, both through experimental activities \cite{wilsonAngularMomentumGeneration2021,GihaGammaMult,Gjestvang, Qi,Laborie,Aumann} and theoretical modeling \cite{KawanoMemoryofSpin,vogtAngularMomentumEffects2021,BulgacSpinEnergy}.

Within the Hauser-Feshbach fission fragment de-excitation model, Kawano \textit{et al.} \cite{KawanoMemoryofSpin} indeed found indications that FFs from photofission and neutron-induced fission might have similar angular momenta at the same excitation energy of the CN, regardless of its spin. 
Similarly, in the FREYA code \cite{vogtAngularMomentumEffects2021}, Randrup and Vogt assume that the main fraction of the CN angular momentum goes into the relative motion of the FFs, while their spin is mainly determined by the temperature of the fissioning nucleus at scission.
Furthermore, within the framework of a fully microscopical model, Bulgac \textit{et al.} showed \cite{BulgacSpinEnergy} that an increase in the compound nucleus excitation energy might produce FFs at higher spin. They concluded that the average spin of the FFs increases with the kinetic energy of the incident neutrons in the fission reactions of $^{235}$U and $^{239}$Pu.  

From an experimental perspective, Bulgac's result is supported by References \cite{GihaGammaMult, Laborie,Qi} which report an increase of the $\gamma$-ray multiplicity as a function of the kinetic energy of the incident neutrons in the fission reactions of $^{238}$U and $^{239}$Pu. 

\begin{figure*}[]
    \centering
    \includegraphics[scale = 0.085]{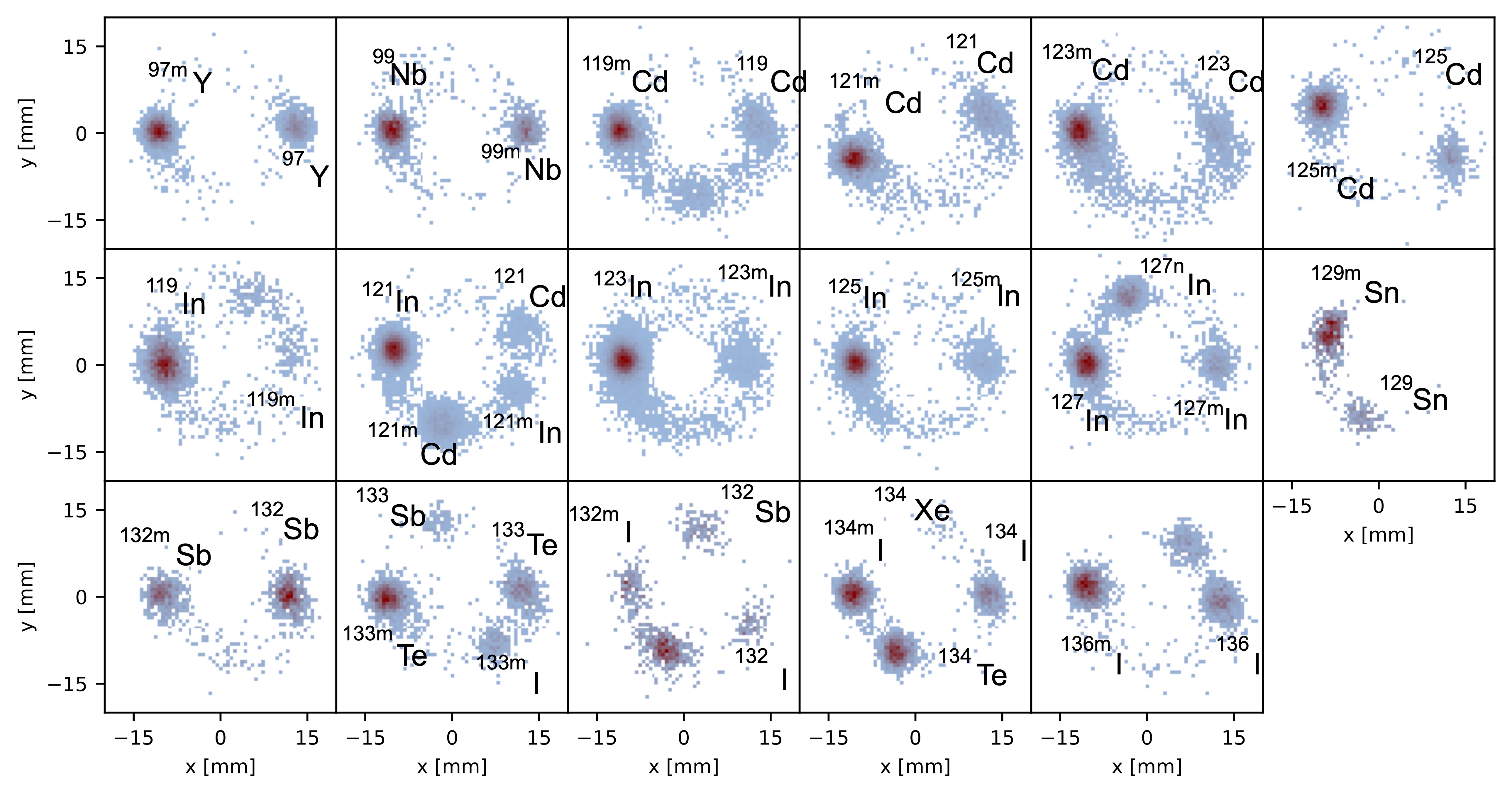}
    \caption{
    PI-ICR images produced during the experimental campaign. Identified nuclei are marked in the plots.
    A number of pictures show the presence of contaminants and asymmetrical tails (see Fig. \ref{fig:polar_proj}). 
    Phase-accumulation times were optimized to ensure that the regions of interest do not overlap with other ion species.
    }
    \label{fig:piicr}
\end{figure*}

A different way of looking for correlations is to use the spin-trap isomers of the produced nuclei. 
The FPs, after the neutron evaporation and the $\gamma$-rays cascade, eventually decay into a (meta)stable nuclear state, before further $\beta$-decay towards the line of stability.
A large number of nuclei have more than one long-lived metastable state, differing in spin and excitation energy.
The angular momentum of a FF influences the population of the isomeric states  \cite{vandenboschIsomericCrossSectionRatios1960,huizengaInterpretationIsomericCrossSection1960}. 
Therefore, the ratio between the production rate of two or more long-lived metastable states, 
the so-called isomeric yield ratio (IYR), provides access to the angular momentum of fission fragments \cite{Wilhelmy,Naik1988,NAIK1995,NAIK2024,rakopoulosFirstIsomericYield2018,Rakopoulos2019,Al-Adili2019,GaoAM}.

\begin{table*}[]
    \centering
    \caption{\label{tab:IYR}
    Measured IYR for fission products from
    $^{232}$Th($\alpha$,f) and relevant nuclear data. The values of half-life (t$_{1/2}$), spin (J), and excitation energy (E$_{\text{exc}}$) were obtained from the Nubase2020 evaluation \cite{NUBASE2020}, except for $^{123}$In and $^{123}$Cd, for which more recent values were retrieved from the ENSDF database \cite{ENSDF}. Values for half-life and spin are given for ground (g) and excited (e) state in the isomeric pair. $^{127}$In has two long-lived excited states; however, in this work, only the ground state and the 394 keV, J=1/2$^-$ state are considered.
    }
    \begin{ruledtabular}
    \begin{tabular}{cccccccc}
Nucleus & t$_{1/2,g}$ [s] & J$_g\;[\hbar]$ & E$_{\text{exc}}$ [keV] & t$_{1/2,e}$ [s] & J$_e\;[\hbar]$ & IYR \\
\colrule
$^{97}$Y & $3.75(3)$ & 1/2- & $667.5(2)$ & $1.17(3)$ & 9/2+ & $0.800(5)$ \\
$^{99}$Nb & $15.0(2)$ & 9/2+ & $365.27(8)$ & $150(10)$ & 1/2- & $0.732(8)$ \\
$^{119}$Cd & $161(1)$ & 1/2+ & $146.5(1)$ & $132(1)$ & 11/2- & $0.836(5)$ \\
$^{121}$Cd & $13.5(3)$ & 3/2+ & $214.9(1)$ & $8.3(8)$ & 11/2- & $0.834(5)$ \\
$^{123}$Cd & $2.1(3)$ & 3/2+ & $144(4)$ & $1.8(3)$ & 11/2- & $0.85(1)$ \\
$^{125}$Cd & $0.68(4)$ & 3/2+ & $186(4)$ & $0.48(3)$ & 11/2- & $0.86(1)$ \\
$^{119}$In & $144(6)$ & 9/2+ & $311.37(3)$ & $1.08(2) \cdot 10^{3}$ & 1/2- & $0.948(3)$ \\
$^{121}$In & $23.1(6)$ & 9/2+ & $313.68(7)$ & $233(6)$ & 1/2- & $0.947(5)$ \\
$^{123}$In & $6.17(5)$ & 9/2+ & $327.21(4)$ & $47.4(8)$ & 1/2- & $0.958(3)$ \\
$^{125}$In & $2.36(4)$ & 9/2+ & $352(12)$ & $12.2(2)$ & 1/2- & $0.933(3)$ \\
$^{127}$In & $1.086(7)$ & 9/2+ & $394(12)$ & $3.62(2)$ & 1/2- & $0.873(3)$ \\
$^{129}$Sn & $134(2)$ & 3/2+ & $35.15(5)$ & $414(6)$ & 11/2- & $0.80(1)$ \\
$^{132}$Sb & $167(4)$ & 4+ & $150(50)$ & $246(3)$ & 8- & $0.41(1)$ \\
$^{133}$Te & $750(20)$ & 3/2+ & $334.26(4)$ & $3.32(2) \cdot 10^{3}$ & 11/2- & $0.708(7)$ \\
$^{132}$I & $8.26(5) \cdot 10^{3}$ & 4+ & $110(11)$ & $4.99(5) \cdot 10^{3}$ & 8- & $0.59(2)$ \\
$^{134}$I & $3.15(1) \cdot 10^{3}$ & 4+ & $316.5(2)$ & $211(2)$ & 8- & $0.715(7)$ \\
$^{136}$I & $83.4(4)$ & 1- & $206(15)$ & $47(1)$ & 6- & $0.692(8)$ \\
\end{tabular}
\end{ruledtabular}
\end{table*}

In this work, we investigate how the CN state affects the angular momentum of FFs by making use of the IYR and the assumption of a positive correlation between the FFs spin and the population of high-spin isomeric states in the observed fission products. 

We define the IYR as
\begin{equation}
  \text{IYR} = \frac{\text{Y}_{\text{hs}}}{\text{Y}_{\text{hs}}+\text{Y}_{\text{ls}}}, 
  \label{eq:IYR_def}
\end{equation}
where Y$_{\text{hs}}$ and Y$_{\text{ls}}$ are the fission product yields of the high-spin and low-spin state for an isomeric pair.

We study comparable fissioning nuclei using both new IYR data from our measurement of the 28 MeV $\alpha$-induced fission of $^{232}$Th (see Table \ref{tab:IYR}) and literature data retrieved from the EXFOR experimental nuclear data library \cite{EXFORExperimentalNuclear}. 
Using 28 MeV $\alpha$-particles, the CN formed by $^{232}$Th($\alpha$,f), \textit{i.e.} $^{236}$U*, may emit one or two pre-scission neutrons.
The probabilities of the first-, second-, and third-chance fission calculated using the nuclear model codes TALYS v2.0 \cite{Koning2023} and GEF v3.3 \cite{GEF}, are listed in Table \ref{tab:gef_talys}, together with the corresponding average excitation energies. 
One can also produce $^{236}$U* and $^{234}$U* via thermal neutrons impinging on $^{235}$U and $^{233}$U, respectively. In both cases the excitation energy is approximately 6 MeV.

\begin{figure}[]
    \centering
    \includegraphics[scale = 0.8]{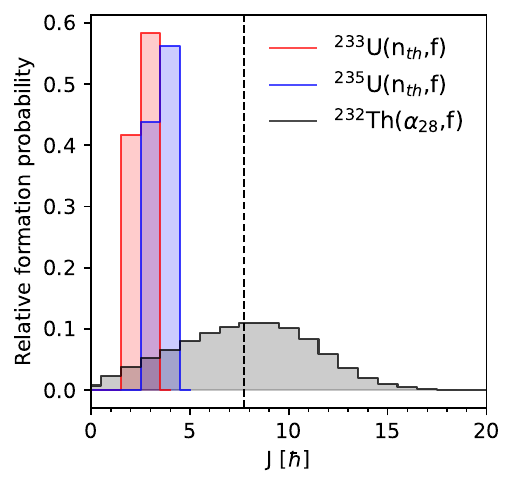}
    \caption{Relative formation probabilities of CN angular momenta resulting from the three considered reactions. The average values are 2.6 $\hbar$ and 3.6 $\hbar$ for $^{234}$U* and $^{236}$U*, respectively, when produced from thermal neutrons, and 7.7 $\hbar$ for $^{236}$U* produced from incident alpha particles at 28 MeV. }
    \label{fig:ang_mom_distr}
\end{figure}

According to optical model calculations with TALYS using default parameters, the CN produced by 28 MeV $\alpha$-particle impinging on $^{232}$Th has an average angular momentum of 7.7 $\hbar$, as shown in Figure \ref{fig:ang_mom_distr}. 
We assume that pre-scission neutrons do not, on average, change the spin of the system, an assumption commonly made in studies of de-exciting fission fragments.
The thermal (s-wave) neutrons produce $^{236}$U* and $^{234}$U* with average angular momenta of about 3.6 $\hbar$ and 2.6 $\hbar$ (Fig. \ref{fig:ang_mom_distr}), respectively.
Therefore, the CN produced in these three fission reactions are similar but carry significantly different angular momenta. 
We assume that these distributions represent the angular momenta of those nuclei that actually fission.
\begin{table}[]
    \centering
\caption{\label{tab:gef_talys}
Probability of first- to third-chance fission for the 28 MeV $\alpha$-induced fission of $^{232}$Th, as calculated by TALYS (P$_{T}$) and GEF (P$_{G}$), and corresponding average excitation energies ($\overline{E}_{T}$, $\overline{E}_{G}$).}
    \begin{ruledtabular}
        
    \begin{tabular}{lllllll}
Scissioning \\ nucleus & P$_{T}$ [\%] & P$_{G}$ [\%] & $\overline{\text{E}}_{T}$ [MeV] & $\overline{\text{E}}_{G}$ [MeV] \\
\colrule
$^{236}$U* & 24 & 21 & 23.0 & 23.0 \\
$^{235}$U* & 35 & 41 & 14.9 & 15.4 \\
$^{234}$U* & 40 & 37 & 8.4 & 8.9 \\
\end{tabular}
\end{ruledtabular}
        
\end{table}

While IYR from thermal-neutron induced fission reactions have been measured for several nuclei, the corresponding data for $\alpha$-induced fission of $^{232}$Th is scarce \cite{Sears2021118} with only one measurement for $^{131}$Te and $^{133}$Te \cite{Datta1983}.
The IYR for $^{133}$Te reported in Ref.~\cite{Datta1983} differs significantly from our result.
This is likely due to limitations in their measurement method, which requires significant corrections based on fission yields and decay data.

We measured 17 IYRs from the $^{232}$Th($\alpha$,f) reaction using the Phase-Imaging Ion-Cyclotron-Resonance (PI-ICR) technique of the JYFLTRAP double Penning trap mass spectrometer, at the IGISOL-4 facility \cite{EliseevPiicr,Moore2013,Al-Adili2019}.
The technique has previously been used to measure IYR \cite{Rakopoulos2019,Gao2023_machine,Gao2023_iyrs}
and is applicable for isomeric states with half-lives longer than approximately 500 ms, and can resolve states separated by 30-40 keV, or more. 
The obtained data are presented in Table \ref{tab:IYR}.
Out of these, 16 have been measured for this system for the first time. 

In the experiment, 32 MeV $\alpha$-particles from the K-130 cyclotron were directed to a $^{232}$Th target with a thickness of 10 mg/cm$^2$. 
The average $\alpha$-particle kinetic energy in the thorium target was estimated to be 28 MeV. 
In the stopping cell of the fission ion guide (see, \textit{e.g.}, \cite{Penttila20121}) primary ions from the fission reactions were stopped and extracted by a purified helium gas flow at 300 mbar. 
Due to the high ionization potential of helium, the fission products (FPs) are not neutralized in this process, allowing them to be accelerated to 30 keV using an electrostatic potential.

\begin{figure}[h]
    \centering
    \includegraphics[scale =.7]{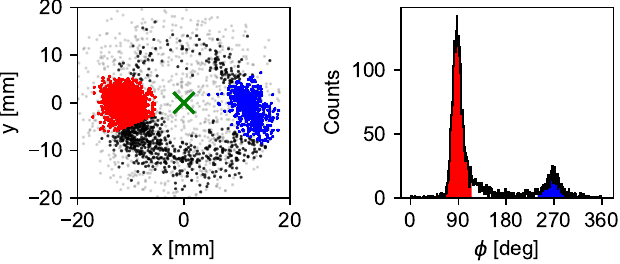}
    \caption{
    PI-ICR image (left) for $^{123}$Cd and polar projection of the detected ions (right). 
    In blue and red the ions identified as high and low spin isomeric states are shown. 
    }
    \label{fig:polar_proj}
\end{figure}

The formed ion beam was mass-separated using a 55$\degree$ dipole magnet with a mass resolution M/$\Delta$M of approximately 500. The continuous beam was collected in a cooler and buncher \cite{Nieminen}, where it was bunched for high-precision mass analysis in the JYFLTRAP double Penning trap \cite{Eronen20121}. 
In a first step, as much of the contaminants as possible were removed from the ion bunch in the first trap using the mass-selective buffer-gas cooling method \cite{SAVARD1991247}. 
In the second trap, the PI-ICR technique \cite{nesterenkoStudyRadialMotion2021} was applied.
First the ion motion was excited with a short (1 ms) dipolar oscillating electric field at the ions' reduced cyclotron frequency in order to initiate the mass-dependent revolution of the ions in the trap. Over time, ions with different mass separate due to their different eigenfrequencies. 
After a pre-set waiting time, a quadrupole excitation at the ions free-space cyclotron frequency 
\mbox{$\nu_c = \frac{1}{2\pi} \frac{q}{m} B$}
is used to convert the fast mass-dependent cyclotron motion to the slow magnetron motion. 
This "freezes" the relative position of the ions, which are extracted and detected in a position-sensitive detector, as illustrated in Figure \ref{fig:piicr} and Figure \ref{fig:polar_proj}.

The two spots are formed by the detected nuclei of interest in the two isomeric states. The method to assign the spots to the corresponding isomeric state is explained in detail in Ref. \cite{rakopoulosFirstIsomericYield2018,Gao2023_machine}. For both spots, the ions detected within 3$\sigma$ from the center of the peaks in the polar projection (right plot in Figure \ref{fig:polar_proj}) are used to calculate the number of counts in the spot.
In some cases, the spots in PI-ICR images show asymmetric tails. 
The physics origin of these is still unclear, but likely due to ion-ion interactions. 
In this measurement, the number of ions in these tails is on the order of 1-5\% of the number of ions in the corresponding spot.
To account for the presence of the tails, different methods to count the detected ions were tested, and in cases where the methods resulted in different values, the uncertainty was increased to enfold all data.  

Each measurement was performed a second time, where the position of the spots on the MCP was interchanged to minimize the corrections for inhomogeneities in the MCP detection efficiency. 
Following the definition of IYR given in Eq. \ref{eq:IYR_def}, we calculate the ratio between the number of observed ions in either state extracted from the trap as
\mbox{$\text{IYR}_{\text{exp}} = \text{C}_{\text{hs}}/\left(\text{C}_{\text{hs}}+\text{C}_{\text{ls}}\right)$},
where C$_{\text{hs}}$ and C$_{\text{ls}}$ are the number of counts associated with the high-spin and low-spin isomeric states, respectively.
The two $\text{IYR}_{\text{exp}}$ were eventually merged into an average value.

\begin{figure}[]
    \centering
    \includegraphics[scale =.8]{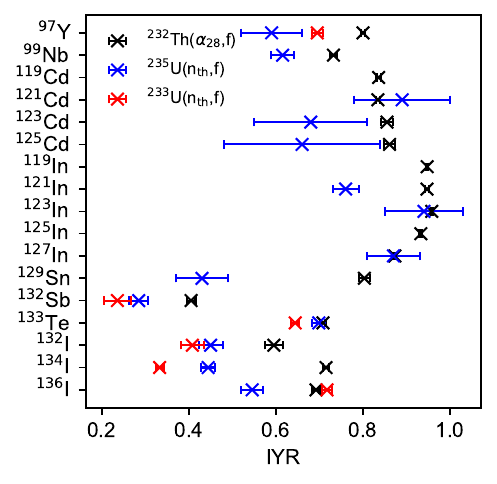}
    \caption{Weighted average value of experimental IYR in literature for thermal neutron-induced fission of $^{233}$U (red) and $^{235}$U (blue) compared to the $^{232}$Th data.}
    \label{fig:IYRs_bib}
\end{figure}
 
If the half-lives of the nuclei of interest or their $\beta$-decay precursors are similar to the measurement cycle (0.5-1 s), a relevant fraction of them will have time to decay before detection. 
The procedure to correct for this started with an assumed value for the isomeric ratio after fission, \textit{i.e.} the true IYR. 
The ion transport was simulated by discretizing the transport time into time steps, and evaluating, for each, a balance equation for the isomeric states of the ion of interest and its precursors. 
This process was iterated, adjusting the trial ratio until the resulting calculated IYR matched the experimental one. 
Bootstrapping is used to determine the IYR and its uncertainty.
Details are presented in Ref. \cite{CanLic}.

\begin{figure*}[]
    \centering
    \includegraphics[scale =.8]{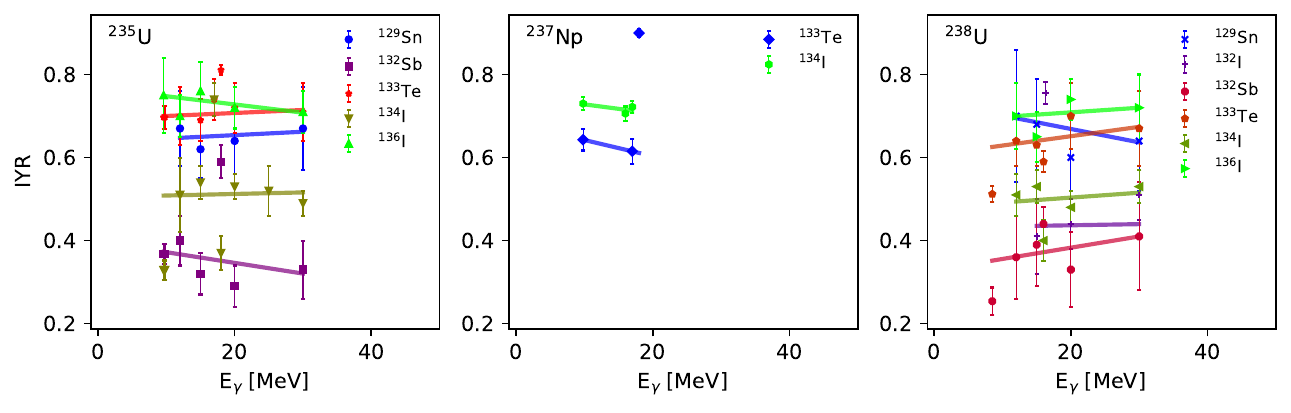}
    \caption{Isomeric yield ratios from photo-fission for the isotopes in this study. The IYR are extracted from EXFOR \cite{EXFORExperimentalNuclear} and plotted as a function of incident photon energy for different fissioning systems. Solid lines represent the linear regressions of the experimental data, where the Huber loss function \cite{HuberRegr} is used in order to have a regression robust to outliers. Only reactions with more than three energy data-points are included.}
    \label{fig:en_dep}
\end{figure*}

The measured IYR from the present study are listed in Table~\ref{tab:IYR} and shown in Figure~\ref{fig:IYRs_bib}. Except for $^{132}$Sb, all nuclei show a dominant feeding of the high-spin state, \textit{i.e.} an IYR larger than 0.5. 
Figure~\ref{fig:IYRs_bib} also shows the comparison between the IYRs measured in this study and those measured in the neutron-induced fission reactions retrieved from literature data, where the EXFOR data are averaged into one data point for each nucleus and fissioning system. 
The $^{232}$Th IYR data consistently show larger values than the $^{233}$U and $^{235}$U data (Figure \ref{fig:IYRs_bib}). 
The weighted means of the difference between experimental and literature data are $\Delta$IYR = 0.135(8) and 0.156(6) for $^{235}$U and $^{233}$U, respectively. 
Normalizing with the change in CN spin as described above (from 7.7 $\hbar$ to 3.6 and 2.6 $\hbar$), we arrive at $\Delta$IYR/$\Delta \overline{\text{J}}$ = 0.032(1)  $\hbar^{-1}$ for the studied isomeric pairs.
The Wilcoxon signed-rank test suggests a significant difference between our measurements and the literature values (p = 9$\cdot$10$^{-4}$) for $^{235}$U, and a strong trend towards a difference (p = 6.25$\cdot$10$^{-2}$) for $^{233}$U.
These data show that the higher spin CNs produced by $^{232}$Th($\alpha$,f) tend to promote more frequent de-excitation to higher spin isomeric states. 

The used literature data have been obtained with techniques different from the one employed in our work, typically using $\gamma$-ray spectroscopy.  
If provided, we use reported independent IYR but chose to include also cumulative IYR in Figure~\ref{fig:IYRs_bib} and the derived numerical values.
We verified that a restriction to independent IYR gives, within the uncertainties, the same results.
Direct ion-counting is a more advanced technique for measuring IYR (see, \textit{e.g.}, \cite{Rakopoulos2019}), and the difference in measurement methodology may explain some of the observed differences.
However, the found difference in $\Delta$IYR for the two thermal-fission cases as given above is 0.02(1), while directly comparing the two cases gives 0.03(1).
Since the difference in average CN spin is 1 $\hbar$, these
values are in line with the proposed trend.
Nevertheless, the reported values can only serve as preliminary estimates of a IYR dependency on the angular momentum of the compound nucleus. 



To isolate and study the influence of the excitation energy of the CN on the angular momentum of FFs, we compare literature data for the IYRs of several FPs produced in photon-induced fission reactions as a function of the energy of the incident photons. 
In fission reactions induced by photons in the few tens of MeV range, the excitation energy of the CN increases as a function of the incident photon energy while its spin remains mainly unchanged.

Figure \ref{fig:en_dep} shows IYR for the nuclei in this study for photon-induced fission reactions, as a function of the incident photon energy for different fissioning systems. 
These bibliographic data are retrieved from EXFOR \cite{EXFORExperimentalNuclear}.
We obtain an average slope of $\Delta IYR/ \Delta E = 0.0001(5) MeV^{-1}$, \textit{i.e.},
we see no effect of the incident photon energy on the IYR.
This implies that the angular momenta of the FFs seem to remain unaffected by an increase in the CN excitation energy. 

Part of the additional excitation energy can be carried away by neutrons before and after scission. 
Therefore, this observation also implies that neutron emission does not, at least not on average, remove a significant amount of angular momentum.
This agrees with the result by Gjestvang \textit{et al.} \cite{Gjestvang}, and the common assumption that statistical neutrons carry away a negligible amount of angular momentum. 
This has, however, recently been questioned by Stetcu \textit{et al.} \cite{StetcuPromptNuetrons} for post-scission neutrons.

To summarize, we provide experimental evidence that an increase of the spin of the compound system undergoing fission leads to an increase in the population of the high-spin state of an isomeric pair.
Based on the literature \cite{vandenboschIsomericCrossSectionRatios1960,huizengaInterpretationIsomericCrossSection1960}, 
we conclude that this is due to an increase in the FF angular momentum. 
While this might, at least in part, also be due to the a change in excitation energy of the fissioning nucleus, our study of IYR from photo-fission suggests that this is not the case.
In fact, the photo-fission data provide no evidence of any influence of the excitation energy on the IYR and, therefore, the FF angular momentum.

If confirmed, this would contradict the results from the modeling of the momentum generation in FREYA \cite{vogtAngularMomentumEffects2021}, but agrees with the theoretical and experimental observations of References \cite{BulgacSpinEnergy,GihaGammaMult,Laborie,Qi,Aumann}. A PI-ICR measurement of IYR from thermal-neutron induced fission of $^{233,235}$U would be highly desirable and can clarify the situation.

In their research, Wilson \textit{et al.} \cite{wilsonAngularMomentumGeneration2021} studied neutron-induced fission of $^{232}$Th and $^{238}$U at about 2 MeV incident energy, as well as spontaneous fission of $^{252}$Cf. 
From their results they concluded that a FFs' angular momentum does not depend on the fissioning system. 
The same lack of dependency is observed by Sears et al. \cite{Sears2021118} in their IYR compilation. They show, for a number of thermal neutron-induced fission reactions, that the measured IYR do not depend on the mass of the fissioning nuclei.  
However, the considered reactions dominantly have small CN angular momenta.
Combining these findings with our results would imply that while a part of the FF angular momentum comes from the fission process itself, it is also influenced by the angular momentum present in the CN. 

Based on our results, one can estimate a lower limit of the additional angular momentum of a FF
that is due to an increase in CN angular momentum.
Using the TALYS code, we have estimated the additional angular momentum that a FF needs to have to increase the observed IYR by 0.15.
We found that, on average, this increase needs to be at least one unit of $\hbar$.
This would imply that at least 2 $\hbar$ out of the additional 4-5 $\hbar$ (\textit{i.e.}, $>$40 \%) goes to the FFs' angular momenta, while a maximum of 3 $\hbar$ (\textit{i.e.}, $<$60\%) is converted into orbital angular momentum or dissipated via emission of pre-fission neutrons.
A detailed calculation could be done using methods outlined in Refs.\cite{rakopoulosFirstIsomericYield2018,GaoAM,Al-Adili2019}.
The values given here should be considered as a conservative estimate of a lower limit. 

Finally, we note that the data may suggest a possible dependency of the observed IYR increase on the FP mass, which may be less pronounced in the region of symmetric fission. 
This could mean that a larger portion of the additional angular momentum is transmitted to the FFs in asymmetric fission compared to symmetric fission. 
To investigate this hypothesis, more data is needed for isotopes in the region around mass number A=116, where several suitable isomers exist. 

\begin{acknowledgements}
 We acknowledge the staff of the Accelerator Laboratory of University of Jyväskylä (JYFL-ACCLAB) for providing a stable online beam.
 We express gratitude for the fruitful discussions with Toshihiko Kawano.
 This work was supported by the Swedish Research Council
 (Ref. No. 2020-04238) and has received funding from the Euratom research and training program 2014-2018 under grant agreement No. 847594 (ARIEL).
 We acknowledge support from the Research Council of Finland under grant numbers 295207, 327629, and 354968, the European Union’s Horizon 2020 Research and Innovation Programme under Grant Agreement No's 771036 (ERC CoG MAIDEN) and 861198-LISA-H2020-MSCA-ITN-2019.
\end{acknowledgements}

\bibliographystyle{apsrev4-2}
\bibliography{main.bib}

\end{document}